\documentclass[aps,prl,superscriptaddress,showpacs,twocolumn,floatfix]{revtex4}
\usepackage{graphicx}
\usepackage[usenames,dvipsnames]{color}
\newcommand{\tc}{,~}

\newcommand{\gevsq}{GeV$^2$}

\newcommand{\qsq}{$Q^2$}

\newcommand{\gmn}{$\mathrm {G_{_{\rm M}}^{\it \,n}}$}
\newcommand{\rea}{$\mathrm{^{3}\overrightarrow {\mathrm{He}}
(\vec{\it e},{\it e'n}){\it pp}}$}
\newcommand{\reb}{$\mathrm{^{3}\overrightarrow {\mathrm{He}}
(\vec{\it e},{\it e'p}){\it np}}$}
\newcommand{\he}{$\mathrm{^{3}He}$}
\newcommand{\Am}{${\rm A_{meas}}$}
\newcommand{\Dt}{${\rm D_{t}}$}
\newcommand{\Db}{${\rm D_{bkgr}}$}
\newcommand{\Ab}{${\rm A_{bkgr}}$}
\newcommand{\Aph}{${\rm A_{phys}}$}
\newcommand{\Din}{${\rm D_{in}}$}
\newcommand{\Ain}{${\rm A_{in}}$}
\newcommand{\Dpn}{${\rm D_{\it p/n}}$}
\newcommand{\AQE}{${\rm A_{_{QE}}}$}
\newcommand{\Aep}{${\rm A_{\it ep}}$}
\newcommand{\Aen}{${\rm A_{\it en}}$}

\newcommand{\GEn}{\mbox{${ G_{_{\!E}}^{\,n}}$}}
\newcommand{\GMn}{\mbox{${ G_{_{\!M}}^{\,n}}$}}
\newcommand{\gn}{\mbox{$\mu_n{ G_{_{\!E}}^{n}}$}/\mbox{${ G_{_{\!M}}^{n}}$}}
\newcommand{\GEp}{\mbox{${ G_{_{\!E}}^{p}}$}}
\newcommand{\GMp}{\mbox{${ G_{_{\!M}}^{p}}$}}
\newcommand{\PR}{{ Phys. Rev. }}

\newcommand{\PRL}{{ Phys. Rev. Lett. }}

\newcommand{\etal}{{\em et al.}}
\newcommand\CMU{Carnegie Mellon University, Pittsburgh, PA 15213}
\newcommand\UBasel{Universit\"{a}t Basel, CH-4056 Basel, Switzerland}
\newcommand\UVa{University of Virginia, Charlottesville, VA 22903}
\newcommand\Yerevan{Yerevan Physics Institute, Yerevan 375036, Armenia}

\newcommand\TelAviv{Tel Aviv University, Tel Aviv, 69978 Israel}
\newcommand\FIU{Florida International University, Miami, FL 33199}

\newcommand\JLab{Thomas Jefferson National Accelerator Facility,
  Newport News, VA 23606 }

\newcommand\NCCU{North Carolina Central University, Durham, NC 27707}
\newcommand\Syra{Syracuse University, Syracuse, NY 13244}
\newcommand\Kent{Kent State University, Kent, OH 44242}
\newcommand\NSU{Norfolk State University, Norfolk, VA 23504}
\newcommand\ODU{Old Dominion University, Norfolk, VA 23529}
\newcommand\GU{University of Glasgow, Glasgow G12 8QQ, Scotland, U.K.}
\newcommand\CalSt{California State University Los Angeles\tc 
Los Angeles, CA 90032}
\newcommand\MIT{Massachusetts Institute of Technology, Cambridge, MA 02139}
\newcommand\BINP{Budker Institute for Nuclear Physics\tc Novosibirsk 630090, 
Russia}
\newcommand\TINP{Institute for Nuclear Physics\tc Tomsk 634050, Russia}
\newcommand\UNH{University of New Hampshire, Durham, NH 03824}
\newcommand\WaM{College of William and Mary, Williamsburg, VA 23187}
\newcommand\Temple{Temple University, Philadelphia, PA 19122}
\newcommand\Kharkov{Kharkov Institute of Physics and Technology\tc Kharkov 
61108, Ukraine}
\newcommand\Duke{Duke University and TUNL, Durham, NC 27708}
\newcommand\ClF{Universit\'{e} Blaise Pascal/IN2P3\tc F-63177 Aubi\`{e}re, 
France}

\newcommand\UMd{University of Maryland, College Park, MD 20742}
\newcommand\INFN{INFN gr. Sanit\`a coll. Sezione di Roma and Istituto 
Superiore di Sanit\`a, Rome, Italy}

\newcommand\RU{Rutgers, The State University of New Jersey\tc Piscataway, 
NJ 08854}
\newcommand\UMASS{University of Massachusetts, Amherst, MA 01003}
\newcommand\Korea{Kyungpook National University, Taegu City, South Korea}
 
\newcommand\Spain{Universidad Complutense de Madrid, Madrid, Spain}
\newcommand\KU{University of Kentucky, Lexington, KY 40506}
\newcommand\ANL{Physics Division, Argonne National Laboratory, Argonne, IL 60439}
\newcommand\IEM{Instituto de Estructura de la Materia, Madrid, Spain}

\newcommand\UIUC{University of Illinois, Urbana-Champaign, IL 61801}

\begin{document}

\preprint{APS/123-QED}

\title{Measurements of the Electric Form Factor of the Neutron \\
up to $Q^{2}=3.4$ GeV$^2$ using the Reaction \rea{}}
\author{S.~Riordan}       \affiliation{\CMU} \affiliation{\UVa} \affiliation{\UMASS}
\author{S.~Abrahamyan}    \affiliation{\Yerevan}
\author{B.~Craver}        \affiliation{\UVa}
\author{A.~Kelleher}      \affiliation{\WaM}
\author{A.~Kolarkar}      \affiliation{\KU}
\author{J.~Miller}        \affiliation{\UMd}
\author{G.D.\ Cates}      \affiliation{\UVa}
\author{N.~Liyanage}      \affiliation{\UVa}
\author{B.~Wojtsekhowski} \thanks{Corresponding author: bogdanw@jlab.org}
\affiliation{\JLab}
\author{A.~Acha}	\affiliation{\FIU}
% ...

\date{\today}

\begin{abstract}
The electric form factor of the neutron was determined from studies 
of the reaction \rea{} in quasi-elastic kinematics in Hall A at Jefferson Lab.
Longitudinally polarized electrons were scattered off a polarized target 
in which the nuclear polarization was oriented perpendicular to 
the momentum transfer.  
The scattered electrons were detected in a magnetic spectrometer in
coincidence with neutrons that were 
registered in a large-solid-angle detector.  
More than doubling the $Q^2$-range over which it is known, we 
find \GEn{}$ = 0.0236 \pm 0.0017 (stat) \pm 0.0026 (syst)$,
$0.0208 \pm 0.0024 \pm 0.0019$, and $0.0147 \pm 0.0020 \pm 0.0014$ for 
$Q^2$ = 1.72, 2.48, and 3.41~\gevsq, respectively. 
\end{abstract}
\maketitle

Understanding the nucleon in terms of QCD degrees of freedom
requires precision measurements of nucleon structure, including the
form factors (FFs) that govern the elastic scattering of electrons.
Important advances in such efforts came 
from the determination, at Jefferson Lab (JLab),
of the ratio of the electric and magnetic elastic FFs of the
proton, \GEp{}/\GMp{}, over a range of the negative four-momentum 
transfer squared (\qsq) of 1 to 6~\gevsq{}~\cite{jo00}.
The ratio \GEp{}/\GMp{} was observed to decrease
almost linearly with increasing \qsq{}, when expectations, based on 
both earlier cross-section measurements and 
prevailing theoretical models of 
the nucleon, had been that such a ratio is constant.
This observation has clarified the necessity for a reconsideration of nucleon
structure with an increased emphasis on the significance of quark orbital 
angular momentum (OAM), see e.g. the review~\cite{boffi07}.
Evidence of quark OAM has subsequently been observed in several 
other independent contexts~\cite{zhe04}.
Given the important implications of Ref.~\cite{jo00}, it is critical to 
determine the neutron form-factor ratio, \GEn/\GMn, in 
a \qsq-region where the unexpected results for the proton were observed, and 
thus to test the theoretical explanations that have emerged for the proton data.

The powerful method of determining FFs using double-polarization 
asymmetries~\cite{ak58}, which led to the striking results of \cite{jo00}, 
has also been used to study $g_n = \mu_n$\GEn/\GMn, 
where $\mu_n = -1.913$ is the neutron magnetic moment, up to \qsq=1.5~\gevsq{}.
These experiments have employed polarized electrons
and either a neutron polarimeter~\cite{gl05, pl06}, a polarized
deuteron target~\cite{zh01, wa04}, or a polarized \he{} 
target~\cite{jo91,be99,go01,be03b}.
At low momentum transfer, the nuclear effects in double-polarization asymmetries
have been taken into account using precise non-relativistic
calculations of \he{} based on the Faddeev-like integral equations~\cite{gl96}, 
whereas at large \qsq{} the eikonal approximation~\cite{gl55} provides 
sufficient precision.
For \qsq-values of several \gevsq{}, even polarization-based studies 
of $g_n$ become very challenging due to the small cross sections involved, 
thus necessitating significant technical development.

We report a measurement of $g_n$, 
up to \qsq=3.4~\gevsq, performed at JLab in experimental Hall A. 
The experiment was made possible through the use of a high-luminosity 
optically-polarized \he~target, a magnetic spectrometer of 76~msr 
solid angle to detect the scattered electrons, 
and a large neutron detector with matched acceptance.
The typical \he-electron luminosity was 
\mbox{$5 \times 10^{35}$ cm$^{-2}$/s}.
The central kinematics, as well as the average values of various
experimental parameters, are listed in Table~\ref{tab:kinematics}.
\begin{table}[!hbt]
\vspace{-0.4 cm}
\caption{Kinematics and other parameters of the experiment: the negative 
four-momentum transfer, \qsq; the rms of \qsq{} range, $\Delta Q^2$;
beam energy, E$_{\rm {beam}}$; central angle of the electron 
spectrometer, $\theta_{\rm e}$; central angle of the neutron detector,
$\theta_{\rm n}$; distance from the target to the neutron detector, D;
longitudinal beam polarization, $P_{\rm e}$; 
target polarization, $P_{_{\rm {He}}}$.} 
\begin{ruledtabular}
\renewcommand{\arraystretch}{1.2}
\begin{tabular}{clccc}
  $\langle Q^2 \rangle $ &[\gevsq]   & 1.72   & 2.48 & 3.41    \\
\hline
  $\Delta Q^2$           &[\gevsq]   & 0.14   & 0.18 & 0.22    \\
%  \hline
  E$_{\rm {beam}}$ &[GeV]    & 2.079 & 2.640 & 3.291  \\
  $\theta_{\rm e}$ &[deg]  & 51.6  & 51.6  & 51.6    \\
  $\theta_{\rm n}$ &[deg]  & 33.8  & 29.2  & 24.9    \\
  D          &[m]    & 8.3     & 11  & 11   \\
  $\langle P_{\rm{e}} \rangle $    &[\%]       & 85.2    & 85.0  & 82.9  \\
  $\langle P_{_{\rm {He}}}\rangle $ &[\%]& 47.0  & 43.9  & 46.2   \\
\end{tabular}
\end{ruledtabular}
\label{tab:kinematics}
\vspace{-0.5 cm}
\end{table}

The experiment, E02-013, used a longitudinally polarized electron beam with
a current of 8~$\mu$A.
The helicity of the beam was pseudo-randomly flipped at a rate of 30~Hz. 
The helicity-correlated charge asymmetry was monitored and kept below 0.01\%.
The beam polarization, monitored continuously by a Compton polarimeter,
and measured several times by a M{\o}ller polarimeter~\cite{al04}, 
was determined with a relative accuracy of 3\%.

The polarized \he{} target, while similar in many respects to the 
target described in Ref.~\cite{al04}, included several important improvements.
The \he{} was polarized by spin-exchange with an optically pumped 
alkali vapor, but unlike earlier targets at JLab, 
the alkali vapor was a mixture of Rb and K~\cite{ha01}, rather than Rb alone. 
This greatly increased the efficiency of spin transfer to the \he{} nuclei, 
resulting in a significantly higher polarization. 
The \he{} gas (at a pressure of $\sim$10~atm), a 1\%~admixture of $ 
\mathrm {N_2}$ and the alkali vapor were contained in a sealed glass cell with two  
chambers.
The electron beam passed through the lower ``target" chamber,
a cylinder $\rm 40\,cm$ in length and $\rm 2\,cm$ in diameter,
where the polarization was monitored every six hours with a relative  
accuracy of 4.7\% using NMR.  The polarization was calibrated in the upper
``pumping" chamber using a technique based on electron paramagnetic  
resonance~\cite{ro98}.
A magnetic field of 25~G was created  in the target area by means of 
a 100~cm gap dipole magnet.
The horizontal direction of the field in the target area, 118$^\circ$ with 
respect to the electron beam, was nearly orthogonal to the momentum-transfer 
vector and was measured to 1~mrad accuracy over the length of the target.
The target cell alignment along the beam was regularly checked by varying 
the size of the electron beam spot. The background from beam-cell interactions 
was estimated using data collected with an empty cell and was found to be negligible.

The scattered electrons were detected in the BigBite spectrometer,
originally used at NIKHEF-K~\cite{la98}.
It consisted of a dipole magnet and a detector stack 
subtending a solid angle of 76~msr for a 40~cm long target.
For this experiment, the detector package was completely rebuilt to 
accommodate an increase in luminosity of $10^5$.
The spectrometer was equipped with 15~planes of high-resolution 
multi-wire drift chambers, a two-layer lead-glass 
calorimeter for triggering and pion rejection, and a scintillator hodoscope 
for event timing information.
BigBite provided a relative momentum resolution of $\sim$1\% 
for electrons with a momentum of 1.5~GeV/$c$, a time resolution of 0.25~ns, 
and an angular resolution 
of 0.3~(0.7)~mrad in the vertical (horizontal) direction.
The \qsq-acceptance was $\sim$10\% of the \qsq-value despite 
the large angular acceptance of BigBite, thanks to its large 5:1 
vertical/horizontal aspect ratio.

The recoiling nucleons were detected in coincidence using a large hadron 
detector, BigHAND, that included two planes of segmented veto counters followed 
by a 2.5~cm lead shield, and then seven layers of neutron counters.
Each neutron-counter layer covered a $\rm 1.7\times4\,m^2$ area and was 
comprised of 25(40) plastic scintillator counters that were 5(10)~cm thick.  
All counters were oriented horizontally except for a set of narrow vertical 
bars that were used to calibrate the horizontal coordinate 
measurement.  
A time-of-flight (ToF) resolution of 0.40~ns was achieved, 
and the coordinate resolution was 5~cm.
The efficiency of each veto plane was found to be 97\%.
The detector was shielded on the target side with 5~cm of lead and 1~cm of iron 
and on all other sides with 5~cm of iron.

The trigger was formed using a 100~ns wide coincidence between the signals 
from BigHAND and BigBite, and required the total energy in the BigHAND 
scintillator counters to be above 25~MeV and the total energy deposited 
in the BigBite calorimeter to be above 500~MeV.
A Monte Carlo of our experiment, that included a modeling of the detector response utilizing
Geant4~\cite{Geant4}, was found to be in good agreement with the detector characteristics 
obtained from the experimental data.

The BigBite spectrometer optics were used to reconstruct the momentum, 
direction, and the reaction vertex of the electrons. 
BigHAND was used to determine the direction and charge of 
the recoiling particle.
Using BigBite, it was also possible to accurately determine the time at 
which the scattering event took place, which in turn provided the start 
time for computing the ToF of the recoil particles arriving in BigHAND,
and hence the momentum, $p_n$, of the recoil nucleon.
The three-momentum transfer, $\vec q$, was used to calculate,
for the recoil nucleon, the missing perpendicular momentum,
$p_{_\perp} \,=\, |(\vec q - \vec p_n)\times \vec q|/|\vec q|$
and the missing parallel momentum,
$p_{_\parallel} \,=\, (\vec q - \vec p_n)\cdot \vec q/|\vec q|$.
The invariant mass of the system comprised of the virtual photon 
and the target nucleon (assumed to be free and at rest), $W$, was calculated as
$W \,=\, \sqrt{ m^2 \,+\, 2 \, m \, (E_i-E_f) - Q^2}$, where
$m$ is the neutron mass, $E_i$ the beam energy, and $E_f$ the energy 
of the detected electron.
The identification of quasi-elastic events was largely accomplished using  
cuts on $\,\,p_{_\perp}$ and $W$.
Additional cuts included $p_{_\parallel}$
and the total mass of the undetected hadrons, $\rm m_{un}$.
See Table~\ref{tab:analysis}.

The measured asymmetry was calculated as:
\vspace{-0.2 cm}
\begin{eqnarray}
{\rm A}_{\rm {meas}}^{p(a)} \,=\, \frac{1}{ P_e \, P_{_{\rm {He}}}} \left 
[ \frac{N_+^{p(a)} \,-\, N_-^{p(a)}}{N_+^{p(a)} \,+\, N_-^{p(a)}} \right ],
\end{eqnarray}
where $N^{p(a)}_h$ is the number of events (normalized to beam charge)
with the target polarization
parallel (anti-parallel) to the vector of the holding magnetic
field, and $h$ is beam helicity. 
A statistically-weighted average of ${\rm A}_{\rm {meas}}^p$ and 
${\rm A}_{\rm {meas}}^a$, \Am, was used in the $g_n$ analysis.
In the case of the elastic scattering of 100\% longitudinally polarized 
electrons off 100\% polarized free neutrons, 
in the one-photon approximation, $g_n$ is related to the double spin 
asymmetry, \Aen, through~\cite{do86} \\
\vspace{-1.1 cm}
\begin{center}
\begin{eqnarray}
{\rm A}_{en} &=& \frac
{-2\sqrt{\tau(\tau+1)} \tan(\theta_e/2) \,\cos\phi^* \sin\theta^* \,(g_n/\mu_n)} 
{ (g_n/\mu_n)^2 \,+\, \tau
\left [ 1 \,+\, 2(1+\tau)\tan^2(\theta_e/2) \right ] } + \nonumber \\
 & & \hspace{-1. cm}
 \frac {-2 \tau \sqrt{1+\tau + (\tau+1)^2 \tan^2(\theta_e/2)} 
\tan(\theta_e/2) \,\cos\theta^*}
{ (g_n/\mu_n)^2 \,+\, \tau
\left [ 1 \,+\, 2(1+\tau)\tan^2(\theta_e/2) \right ] } ,
\end{eqnarray}
\end{center}
where $\tau = Q^2/4m^2$, $\theta^*$ is the angle between the  
neutron polarization vector, $\vec P_n$, and $\vec q$, and $\phi^*$ is 
the angle between the electron scattering plane and 
the ($\vec P_n, \, \vec q$) plane.

To obtain $g_n$ from \Am{} a number of corrections were applied, 
the most important of which are presented in Table~\ref{tab:analysis}.
\begin{table}[b]
\vspace{-0.50 cm}
\caption{Data analysis parameters and the resulting asymmetry 
values used to calculate $g_n$ (see text for details).}
\begin{ruledtabular}
\renewcommand{\arraystretch}{1.2}
\begin{tabular}{cccc}
$\langle Q^2 \rangle$ [\gevsq] & 1.72         & 2.48     & 3.41     \\
\hline
  $W$      [GeV]               & 0.7-1.15     & 0.65-1.15 & 0.6-1.15 \\
  $p_{_\perp}$  [GeV]          & $< 0.15$     & $< 0.15$ & $< 0.15$ \\
$p_{_\parallel}$  [GeV]        & $< 0.25$     & $< 0.25$ & $< 0.40$ \\
$\rm m_{un}$  [GeV]            & $< 2.0$      & $< 2.0$  & $< 2.2 $ \\
\hline
  \Am          & -0.136    & -0.134    & -0.098   \\
\hline
  \Dt          &  0.948    &  0.949     &  0.924    \\
  \Db          &  0.970    &  0.981     &  0.975    \\
  \Ab          & -0.001    & -0.018     & -0.012    \\
\hline
  \Aph         & -0.148    & -0.145     & -0.109   \\
\hline
  \Din         &  0.980    &  0.963     &  0.851    \\
  \Ain         & -0.108    & -0.254     & -0.113    \\
\hline
  \AQE         & -0.149    & -0.141     & -0.109    \\
\hline
  \Dpn         &  0.782    &  0.797     &  0.807    \\
$\delta$\Dpn& 0.022 &  0.033 &  0.042 \\
  \Aep         & -0.010    & -0.008     & -0.006   \\
\hline
  $\rm A_{\it en}|_{exp}$ & -0.188    & -0.175    & -0.134   \\ 
\end{tabular}
\end{ruledtabular}
\label{tab:analysis}
\vspace{ -0.20 cm}
\end{table}
A target dilution factor, \Dt, was applied to account for scattering
from the $\mathrm {N_2}$ admixture in the target gas.
Accidental coincidences were accounted for using a background dilution
\Db{} associated with an asymmetry \Ab{} and were determined by
considering the interval of the ToF spectrum that was free from real  
coincidence events.
The resulting physical asymmetry, \Aph,
was then corrected  for inelastic single-pion electroproduction events,
leading to the asymmetry for quasi-elastic processes, \AQE.
The dilution from inelastic events, \Din, and the associated asymmetry,
\Ain, were calculated using our Monte Carlo, which employed the 
plane-wave impulse approximation (PWIA) along with the MAID 
parameterization~\cite{maid}.
The event yield in the Monte Carlo was normalized 
to match the data.
In spite of its significant size, the inelastic background leads to  
only a small correction thanks to the observed asymmetry \Aph{} being close  to \Ain.
The asymmetry $\rm A_{\it en}|_{exp}$ was obtained from \AQE{} using
the dilution factor \Dpn{} and the asymmetry \Aep{} that accounted
for the dilution in our final event sample from protons. 
This dilution was largely due to charge-exchange proton interactions 
in the shielding upstream of the veto planes. 
\Dpn{} and its uncertainty $\delta$\Dpn{} were computed by comparing data collected
from three targets ($\mathrm {H_2}$, \he{}, and $\mathrm {N_2}$).
The asymmetry \Aep{} was computed using the GEA calculations 
in a separate Monte Carlo, as discussed below.

The final steps in extracting $g_n$ involve calculations of the asymmetries 
in the quasi-elastic processes \rea~and \reb.
These calculations were performed using the generalized eikonal 
approximation (GEA)~\cite{sa05},
and included the spin-dependent final-state interactions and 
meson-exchange currents, and used the \he{} wave function 
that results from the AV18 potential~\cite{wi95}.
The yield of the quasi-elastic events and the asymmetries were
calculated as a function of W and assumed values for $g_n$ with
the values for the other nucleon FFs from~\cite{ke04}.
The estimated accuracy of the GEA calculations is 2\%~\cite{sa10}. 
The acceptance of the experimental setup, orientation of the target 
polarization, and the cuts applied to $p_\perp$ and $p_\parallel$ were 
all taken into account.
We note that the effective neutron polarization for the cuts used on 
$p_\perp$ and $p_\parallel$, as calculated in the PWIA approximation, 
was greater than $\sim$96\% of $\rm P_{_{He}}$  
(in agreement with~\cite{ca98}).
The asymmetries for \rea~calculated within GEA were found 
to be within 3\% of the PWIA values, 
indicating that nuclear re-scattering effects were quite small.
The experimental value of $g_n$ and its statistical uncertainty were
calculated by comparing $\rm A_{\it en}|_{exp}$ with the asymmetries 
from the GEA calculations~\cite{sa10}.
The systematic uncertainty was obtained by combining in quadrature the  
contributions of individual effects, as presented in Table~\ref{tab:results}.

\begin{figure}[!t]
\includegraphics[trim = 11mm 10mm 4mm 6mm, height=0.50\textwidth, 
angle = 90]{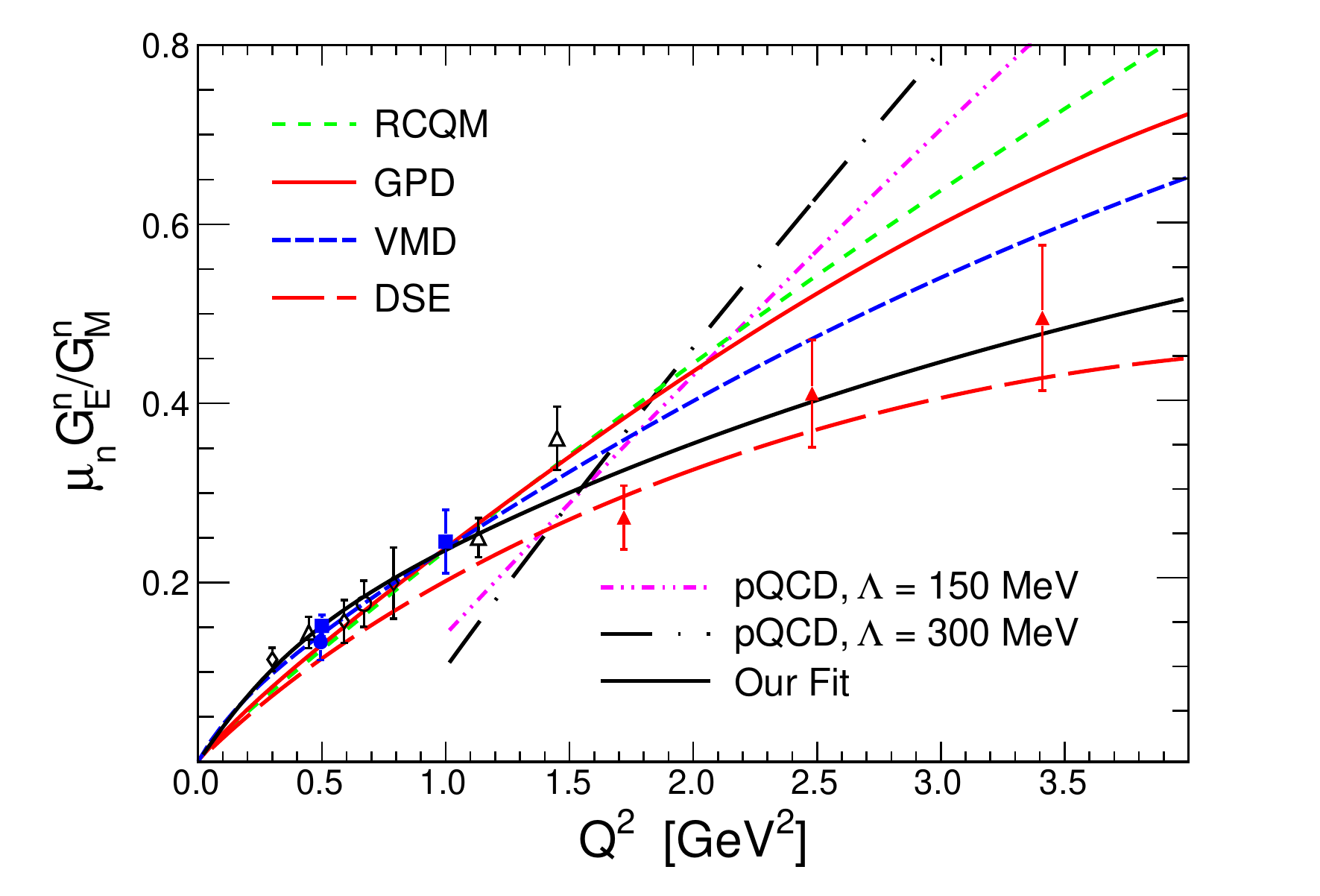}
\caption{The ratio of \gn{} vs. the momentum transfer 
with results of this experiment (solid triangles) and 
selected published data: diamonds~\cite{gl05}, open triangles~\cite{pl06},
circles~\cite{zh01}, squares~\cite{wa04}, open circles~\cite{be03b},
and calculations: pQCD~\cite{be03a}, RCQM~\cite{mi02}, 
DSE~\cite{ro07}, GPD~\cite{di05}, and VMD~\cite{lo02}.
The curves labeled pQCD present pQCD-based scaling 
prediction~\cite{be03a} normalized to 0.3 at \qsq=1.5~\gevsq. 
The error bars for our data points show the statistical and 
the systematic uncertainties added in quadrature.
Our fit is also shown; see parameterization in the text.}
\label{fig:gen}
\vspace{-0.70 cm}
\end{figure}

Our results for $g_n$  are shown in Fig.~\ref{fig:gen} along 
with recent data sets that extend beyond 
\qsq=0.5~\gevsq{}~\cite{gl05, pl06, zh01, wa04, be03b}. 
It is important to compare our results with calculations that have 
described well the proton FF data.
Three such calculations are shown in Fig.~\ref{fig:gen}. 
In all of them, quark orbital angular momentum plays an important role.
One is a logarithmic scaling prediction for the ratio
of the Pauli and Dirac nucleon form factors:
$F_2/F_1 \propto \ln ^2(Q^2/\Lambda^2)/Q^2$~\cite{be03a}, 
based on pQCD, which is shown for two values of the soft-scale 
parameter $\Lambda$. 
It is in clear disagreement with the combined neutron data, 
despite providing a good description
of the proton data.
The authors of \cite{be03a} noted, however, that the agreement with 
the proton data may well have been due to delicate cancellations, 
given the relatively low values of \qsq~ involved.  
Another calculation is the Light Front Cloudy Bag Model~\cite{mi02}, 
an example of a relativistic constituent quark model (RCQM)
calculation that, in this case, includes a pion cloud.
Several RCQMs anticipated the observed decreasing \qsq{}
dependence of \GEp/\GMp.
Finally, we show a calculation based on QCD's Dyson-Schwinger 
equations (DSE)~\cite{ro07}, 
in which the mass of the quark propagators is dynamically generated.
The calculation~\cite{ro07} is closest to our results.
Also shown in Fig.~\ref{fig:gen} are predictions based on 
GPDs~\cite{di05} and Vector Meson Dominance~\cite{lo02}
that were fit to the data available prior to this work.
Finally, our Galster-like fit to the 13 data points (used in
Fig.~\ref{fig:gen}) is shown by a solid black line:
$\mathrm{g_n= \mu_n \left [ a \tau/(1+b\tau) \right ] G_{_D} /}$\gmn, 
where $\mathrm {G_{_D} = 1/(1+Q^2/(0.71 \,  GeV^2))^2}$,  
\gmn{} is from~\cite{ke04}, and we find $a=1.39$, 
$b=2.00$, and a total $\chi^2=7.8$.

\begin{table*}[!ht]
\caption{\label{tab:results} {Experimental results 
for $g_n\equiv \mu_n$\GEn/\GMn{} and \GEn{} 
(using linearly interpolated values of \GMn{} from~\cite{la08}), and 
also the contributions to the systematic uncertainty 
of \GEn{} from individual sources (as a fraction of the \GEn{} value).}}
\begin{tabular}{ccc||ccccccc}
\hline \hline
$ \langle Q^{2} \rangle$ [\gevsq] 
& $\,\,\,\,\,$ $g_n$ $\pm$ stat. $\pm$ syst. $\,\,\,$
& \GEn $\pm$ stat. $\pm$ syst. $\,\,\,$ 
&  $\,\,$ \GMn       $\,\,\,\,\,$
&  $\,$ $P_{_{\rm {He}}}$ $\,\,\,\,$
&  $\,$ $P_n$ $\,\,\,\,$
&  $P_{\rm e} \,\,\,\,$
&  \Dpn $\,\,\,\,$
&  \Din $\,\,\,\,$
&  $\,\, other$ \\
\hline
1.72
&  $0.273 \pm 0.020  \pm 0.030$&$\,\,\,$ $0.0236 \pm 0.0017 \pm 0.0026$ 
$\,\,\,$
& $\,$ 0.020 $\,\,\,\,$ & 0.076 $\,\,\,\,$ & 0.033 $\,\,\,\,$ 
&  0.055 $\,\,\,\,$ & 0.033 $\,\,\,\,$ 
& 0.011 $\,\,\,\,$ & 0.025  \\
2.48
&  $0.412 \pm 0.048  \pm 0.036$&$\,\,\,$ $0.0208 \pm 0.0024 \pm 0.0019$ 
$\,\,\,$ 
& $\,$ 0.024 $\,\,\,\,$ & 0.059 $\,\,\,\,$ & 0.024 $\,\,\,\,$ 
&  0.031 $\,\,\,\,$ & 0.036 $\,\,\,\,$
&  0.027 $\,\,\,\,$ & 0.023  \\
3.41 
&  $0.496 \pm 0.067  \pm 0.046$&$\,\,\,$ $0.0147 \pm 0.0020 \pm 0.0014$ 
$\,\,\,$
&  $\,$ 0.026 $\,\,\,\,$ & 0.047 $\,\,\,\,$ & 0.016 $\,\,\,\,$ 
&  0.026 $\,\,\,\,$ & 0.032 $\,\,\,\,$
&  0.060 $\,\,\,\,$ & 0.026  \\
\hline \hline
\end{tabular}
\label{tab:results}
\vspace{-0.20 cm}
\end{table*}

\begin{figure}[!hbt]
\includegraphics[trim = 6mm 5mm 5mm 5mm, width=0.337\textwidth, angle =
90]{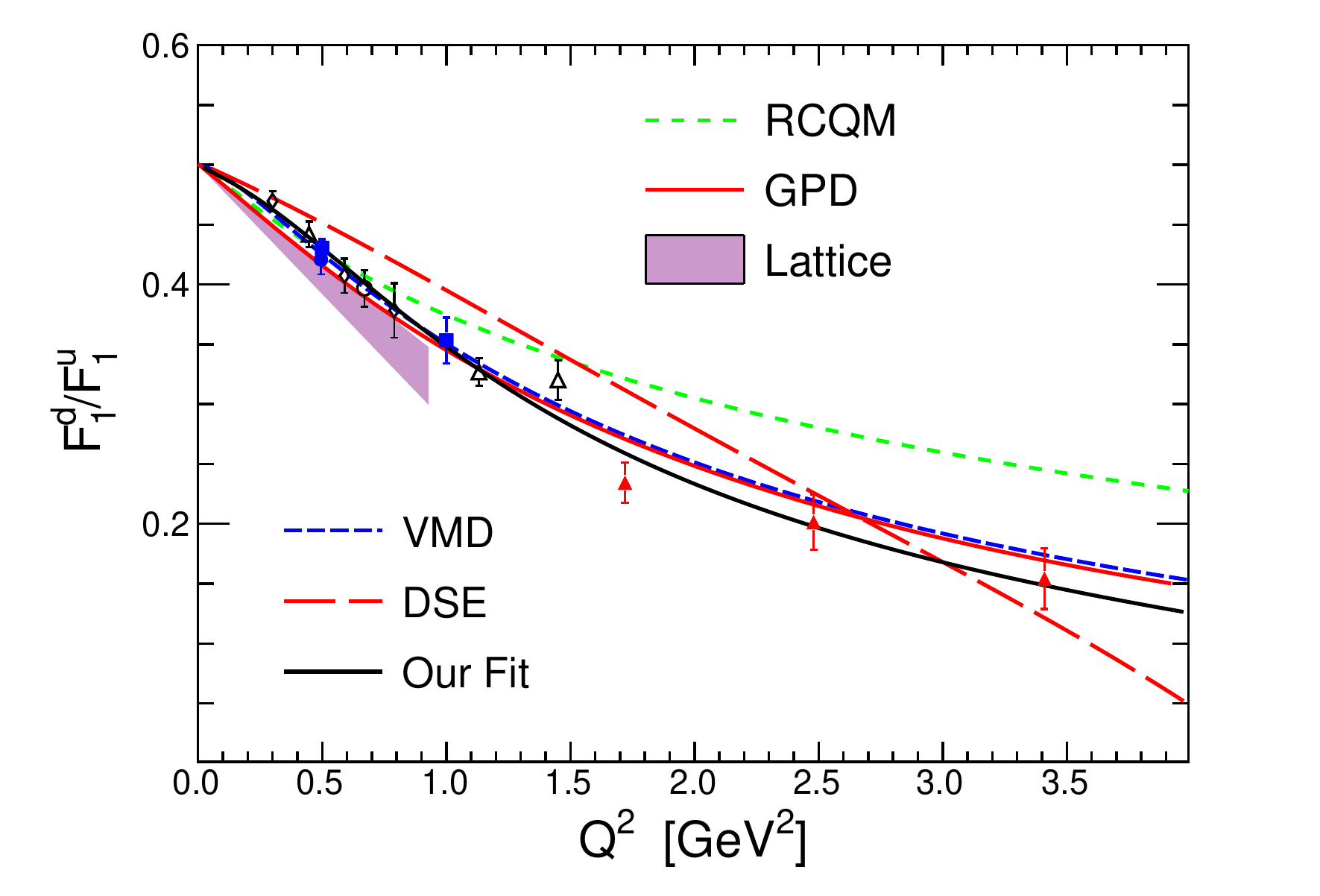}
\caption{Nucleon flavor FF ratio ${F_1^d/F_1^u}$
% and 
%${\kappa_d^{-1} F_2^d/\kappa_u^{-1} F_2^u}$ 
vs \qsq. The band indicates the lattice QCD result~\cite{br10}.
The data and curves correspond to those shown in Fig.~\ref{fig:gen}.
See text for details.} 
\vspace{-0.7cm}
\label{fig:f1df1u} 
\end{figure}

Flavor-separated Dirac and Pauli FFs of the nucleon,
${F_{1,2}^d}$ and ${F_{1,2}^u}$ (for u and d in the proton), can be obtained from 
the electric and magnetic FFs of the proton and the neutron, 
assuming isospin symmetry and neglecting the contribution of 
the strange quark FFs~\cite{mi90}.
Experimental data for $g_n$ and the Kelly fit~\cite{ke04} for \GEp, \GMp{}, and \GMn{} 
were used to compute the ratio ${F_1^d/F_1^u}$, shown in
Fig.~\ref{fig:f1df1u}, which exhibits a downward trend with increasing \qsq{}.
This means that the corresponding infinite-momentum-frame charge 
density~\cite{mi07} 
of the $d$ quark as a function of impact parameter is significantly broader 
than that of the $u$ quarks.
Such an experimental result could be related to the established decrease 
of the quark PDF ratio, $d/u$, with increasing $x_{_{Bj}}$.
The calculations discussed earlier, as well as the recent lattice 
QCD results~\cite{br10}, are in general
agreement with the experimental data for ${F_1^d/F_1^u}$.

We conclude by summarizing in Table~\ref{tab:results} our experimental results.
This experiment more than doubles the \qsq-range over which \GEn{} is known,
greatly sharpens the mapping of the nucleon's constituents and provides
a new benchmark for comparison with theory.
 
\begin{acknowledgments}
We thank the Jefferson Lab Hall A technical staff for their outstanding
support.  This work was supported in part by the National Science
Foundation, the U.S. Department of Energy and the UK Engineering and Physical 
Science Research Council. 
Jefferson Science Associates, LLC, operates Jefferson Lab for 
the U.S. DOE under U.S. DOE contract DE-AC05-060R23177.
\end{acknowledgments}

\end{document}